# Multipulse Double-Quantum Magnetometry With Near-Surface Nitrogen Vacancy Centers

H. J. Mamin,[1] M. H. Sherwood,[1] M. Kim,[1,2] C. T. Rettner,[1] K. Ohno,[3] D. D. Awschalom,[3,4] and D. Rugar[1]

[1]*IBM Research Division, Almaden Research Center, 650 Harry Rd., San Jose CA 95120*
[2]*Center for Probing the Nanoscale, Stanford University, Stanford, CA 94305*
[3]*Center for Spintronics and Quantum Computation, University of California, Santa Barbara CA 93106*
[4]*Institute for Molecular Engineering, University of Chicago, IL 60637*

We discuss multipulse magnetometry that exploits all three magnetic sublevels of the S=1 nitrogen-vacancy center in diamond to achieve enhanced magnetic field sensitivity. Based on dual frequency microwave pulsing, the scheme is twice as sensitive to ac magnetic fields as conventional two-level magnetometry. We derive the spin evolution operator for dual frequency microwave excitation and show its effectiveness for double-quantum state swaps. Using multipulse sequences of up to 128 pulses under optimized conditions, we show enhancement of the SNR by up to a factor of 2 in detecting NMR statistical signals, with a 4× enhancement theoretically possible.

*H. Jonathon Mamin (corresponding author)
mamin@us.ibm.com
408-927-2502

Nitrogen-vacancy (NV) centers in diamond are remarkable quantum mechanical objects that are being widely explored for use in quantum information and nanoscale magnetometry [1–3]. The negatively-charged NV center has an intrinsic S=1 spin state that can be interrogated and reset optically and can have strikingly long coherence time, the key to its sensitivity. One proposed application is to use a near-surface NV center as a sensitive, atomic-size magnetometer to perform magnetic resonance imaging (MRI) at the nanoscale, with possible application to molecular structure imaging [4,5]. The detection of nuclear spins [6–9] in organic samples applied onto a diamond surface has been demonstrated as a first step, as has three-dimensional imaging of near-surface electron spins [10]. Molecular-scale MRI presents significant challenges in signal-to-noise (SNR) ratio due to the weakness of the fields produced by nanoscale ensembles of nuclear spins. Since three dimensional images contain so many pixels, even modest enhancements in SNR can greatly improve data acquisition time.

In most NV magnetometry experiments, the NV center is used as a quasi-two level system, where a single microwave frequency is used to manipulate two of the three triplet-state sublevels. Recently, however, schemes have been implemented which take full advantage of the S=1 nature. With suitable double quantum (DQ) pulse sequences, for example, Toyli *et al.* [11] and others [12,13] demonstrated that the NV can be used to perform sensitive thermometry, while being insensitive to magnetic fields. The opposite regime, in which the NV center is made insensitive to temperature but with enhanced magnetic sensitivity, has also been explored [14–16]. In the present work, we perform DQ magnetometry with multipulse sequences having two simultaneous frequency components. Using optimized DQ sequences with up to 128 pulses, we were able to demonstrate detection of proton spin signals from an external organic layer using a near-surface NV center. A SNR improvement of at least 2× was obtained under optimized conditions. For statistically polarized spin signals [17], a theoretical enhancement of up to 4× may be possible, which would reduce data acquisition time by up to a factor of 16, depending on the NV coherence times.

The basic principle of DQ magnetometry is to make use of the two energy levels corresponding to $m_s = \pm 1$ of the triplet ground state (Fig. 1a) [14–16]. The NV spin Hamiltonian in an applied magnetic field $B$ is $H_0 = \hbar \gamma B S_z + \hbar D S_z^2$, where $D = 2.87$ GHz is the zero-field splitting, $S_z$ is the spin-1 z-operator, and $\gamma/2\pi = 28$ GHz/T is the NV gyromagnetic ratio. The solution to the Schrodinger equation shows that the upper level ($m_s = +1$ state) accumulates phase at a rate given by $D + \gamma B$, while the $m_s = -1$ state does so at a rate given by $D - \gamma B$. Thus a superposition state of $m_s = +1$ and $m_s = -1$ will develop a relative phase that is twice as sensitive to $B$ compared to a single quantum (SQ) superposition with the $m_s = 0$ state. It is also insensitive to small variations in $D$. Since $D$ shifts with temperature by -74 kHz/K (equivalent to roughly 3 µT/K) [18], DQ magnetometry should be more robust against thermal effects.

For ac magnetic fields, the relative phases of $m_s = +1$ and $m_s = -1$ states can be measured using the equivalent of a spin echo or, more generally, a dynamic decoupling sequence [19] (Fig. 1b). Starting with the NV center in the $m_s = 0$ state, a superposition is set up between the $m_s = \pm 1$ states. Periodically throughout the sequence, the state amplitudes are "swapped," in analogy with the pi pulses in a conventional SQ dynamic decoupling sequence. Finally, a "read" pulse is used which, absent any ac field, puts the spin back into the $m_s = 0$ state. Since microwave fields do not cross-couple the $m_s = \pm 1$ states directly, simple monochromatic microwave pulses cannot be used to generate the initial superposition or create the swaps, except at low bias field where the two states are nearly degenerate [14]. For the non-degenerate case, DQ magnetometry experiments typically use a composite consisting of three pulses, each at a single frequency, to perform an echo [15,16]. This has proven to be effective, but requires separate pulsed generators, and adds timing overhead, which may be detrimental in long and rapid multipulse sequences.

In the present work, we replace the composite pulses with single, dual frequency pulses, enabling effective multipulse sequences of the type shown in Fig. 1(b). To understand the evolution of the NV center under the application of a microwave field with two frequency components, we consider the generalized Hamiltonian $H = \gamma B S_z + D S_z^2 + \{\gamma B_1 \cos(\omega_1 t + \delta_1) + \gamma B_2 \cos(\omega_2 t + \delta_2)\} S_x$, where $S_x$ and $S_z$ are the appropriate spin-1 operators, and $B_1$, $B_2$ and $\delta_1$, $\delta_2$ are the amplitudes and phases of the microwave fields applied at frequencies $\omega_1$ and $\omega_2$, respectively ($\hbar \equiv 1$). We apply a time-dependent unitary transformation to $H$, equivalent to operating in the rotating frame for each transition, and drop the rapidly oscillating terms (rotating wave approximation) to produce the transformed Hamiltonian [20]

$$\tilde{H} = \begin{pmatrix} D+\gamma B - \omega_1 & \dfrac{B_1 e^{-i\delta_1}}{2\sqrt{2}} & 0 \\ \dfrac{B_1 e^{i\delta_1}}{2\sqrt{2}} & 0 & \dfrac{B_2 e^{i\delta_2}}{2\sqrt{2}} \\ 0 & \dfrac{B_2 e^{-i\delta_2}}{2\sqrt{2}} & D-\gamma B - \omega_2 \end{pmatrix}.$$

We note that this is a slightly more general version of the Hamiltonian previously considered by Xu *et al.* [21]. In the case of on-resonance irradiation, ($\omega_1 = D+\gamma B$ and $\omega_2 = D-\gamma B$), the matrix can be diagonalized analytically and used to calculate the evolution operator in the transformed frame:

$$U(t,B_1,B_2,\delta_1,\delta_2) = \begin{pmatrix} \dfrac{B_2^2 + B_1^2 \cos(\omega_e t)}{B_1^2 + B_2^2} & \dfrac{-iB_1 e^{-i\delta_1}\sin(\omega_e t)}{\sqrt{B_1^2 + B_2^2}} & \dfrac{(-1+\cos(\omega_e t))B_1 B_2 e^{-i(\delta_1-\delta_2)}}{B_1^2 + B_2^2} \\ \dfrac{-iB_1 e^{i\delta_1}\sin(\omega_e t)}{\sqrt{B_1^2 + B_2^2}} & \cos(\omega_e t) & \dfrac{-iB_2 e^{i\delta_2}\sin(\omega_e t)}{\sqrt{B_1^2 + B_2^2}} \\ \dfrac{(-1+\cos(\omega_e t))B_1 B_2 e^{i(\delta_1-\delta_2)}}{B_1^2 + B_2^2} & \dfrac{-iB_2 e^{-i\delta_2}\sin(\omega_e t)}{\sqrt{B_1^2 + B_2^2}} & \dfrac{B_1^2 + B_2^2 \cos(\omega_e t)}{B_1^2 + B_2^2} \end{pmatrix}$$

where $\omega_e = \gamma\sqrt{B_1^2 + B_2^2}/2\sqrt{2}$.

In the general case of off-resonance irradiation, no simple analytic solution was found, but the matrix could still be diagonalized numerically to give the evolution operator for any particular set of parameters. Both on- and off-resonance calculations were in excellent agreement with numerical simulations performed by integrating the time-dependent Schrodinger equation using the Runge-Kutta method (Fig. S1) [20].

Note that in the particular case where $B_1 = B_2$ and $\delta_1 = \delta_2 = 0$, the evolution operator reduces to

$$U_0(\omega_e t) = \begin{pmatrix} \dfrac{1+\cos(\omega_e t)}{2} & \dfrac{-i\sin(\omega_e t)}{\sqrt{2}} & \dfrac{-1+\cos(\omega_e t)}{2} \\ \dfrac{-i\sin(\omega_e t)}{\sqrt{2}} & \cos(\omega_e t) & \dfrac{-i\sin(\omega_e t)}{\sqrt{2}} \\ \dfrac{-1+\cos(\omega_e t)}{2} & \dfrac{-i\sin(\omega_e t)}{\sqrt{2}} & \dfrac{1+\cos(\omega_e t)}{2} \end{pmatrix},$$ which is simply the standard

spin-1 rotation operator about the $x$-axis [22]. (Similarly, for $\delta_1 = -\delta_2 = 90°$, the operator becomes the spin-1 rotation operator about the y-axis). Thus, in this limit, the effect of a dual-frequency pulse is isomorphic to that of a single on-resonance pulse for a $S=1$ system that has no zero-field splitting ($D=0$). In this limit, setting

$\omega_e t = \gamma B_1 t / 2 = \pi$ gives $U_0(\pi) = \begin{pmatrix} 0 & 0 & -1 \\ 0 & -1 & 0 \\ -1 & 0 & 0 \end{pmatrix}$, which illustrates the ability of a single

DQ pulse to swap the amplitudes of the $m_s = +1$ and -1 states. Examples of the time evolution of different operations are shown in Fig. 1(c) and (d). Depending on the details, an effective swap between the $m_s = \pm 1$ states can occur by first depleting and then replenishing the two states (Fig. 1(c)) or by simultaneously doing both (Fig. 1(d)).

The evolution operators allow for efficient numerical simulation of multipulse sequences, which can help guide the choice of the phases $\delta_1$ and $\delta_2$ in a way that provides some compensation against pulse errors. We consider cases where $\delta_1 = -\delta_2$, appropriate to our method of generating the dual frequencies, which is described below. As an example, it is easily verified that two on-resonance DQ pulses where the phases differ by $\pi$ should exactly cancel each other out; i.e.

$U(t, B_1, B_2, \delta_1, -\delta_1) \cdot U(t, B_1, B_2, \delta_1 + \pi, -(\delta_1 + \pi)) \equiv \mathbf{I}$, the identity matrix. Thus, a multipulse sequence that alternates the microwave phase $\delta_1$ between 0 and 180° should be robust against pulse width errors. Similarly, it can be shown that sequences of multiple DQ swap pulses that alternate in phase between 0 and 90° are forgiving of pulse width errors to 2nd order. Moreover, such sequences are more analogous to conventional XY sequences, which are known to compensate for errors in all three magnetization components [23]. The ability to calculate the time evolution of the spin state through an entire multipulse sequence, including the application of an ac magnetic field, proved valuable in guiding our design of the pulse sequences, which were then tested empirically.

To generate the individual dual frequency swap pulses, a continuous wave signal was used as a local oscillator and heterodyned with a pulsed microwave signal using a double-balanced mixer, thus creating frequency components at the sum and difference frequencies of the two inputs (Fig. 1e) [20,21]. To properly equalize the amplitudes, we separated the signal with a diplexer filter, added the proper attenuation on the individual channels, and then recombined them with another diplexer before sending the microwaves through an amplifer to a microwire that was lithographically defined on the diamond. All pulses and IQ modulation were generated on one RF channel. This approach avoided any issues with synchronization, as both frequency components were automatically included in every pulse.

Our experiments were performed using an electronic-grade single crystal diamond onto which a 64-nm thick layer of isotopically pure $^{12}C$ was grown epitaxially [24]. NV centers were produced approximately 12 nm below the surface by introducing $^{15}N_2$ gas at the appropriate time during growth. The sample was then subjected to $^{12}C$ ion implantation to create vacancies [25], followed by annealing and cleaning.

To demonstrate the difference between SQ and DQ magnetometry, we first used a Hahn echo pulse scheme while applying an ac magnetic test signal synchronized with the echo pulses. Figure 2 shows the echo response plotted as a function of the amplitude of the test signal. The total echo time (40 µs) and pulse widths (32 ns) were identical in the two data sets, as were the data acquisition times and ac field strength; the only difference was the use of DQ rather than SQ pulses. The difference in the periodicity of roughly a factor of 2 is apparent, confirming that the DQ echo is in fact accumulating phase twice as quickly. A reduction in contrast is also apparent. This is a consequence of the shorter $T_2$ for the DQ scheme with a Hahn echo (Fig. 2(c)). This is to be expected if the source of decoherence is Markovian magnetic field noise [16], though other types of decoherence can in some cases be suppressed [26].

Next, we considered the performance of SQ and DQ magnetometry using multipulse sequences. Such sequences provide a convenient way to detect nuclear spins by selectively coupling to magnetic signals at the Larmor frequency of the nuclei [7]. Typically the Larmor frequencies are in the megahertz range; many pulse repetitions are therefore necessary to accumulate sufficient phases for a detectable signal. Because of the numerous pulses, it is desirable to use a sequence that is tolerant to various types of pulse errors. This is especially true in DQ magnetometry, where there are twice as many pulse parameters to optimize.

Figure 3(a) and (b) shows the NV center spin echo response to an ac test magnetic signal using a 64 pulse SQ and DQ sequence respectively, where the total sequence time was 34 µs, and the pulses were carefully chosen for optimum performance [20]. As with the Hahn echo, the period of the DQ curve is roughly half that of the SQ curve. The contrast (amplitude) of the SQ and DQ curves in Fig. 3(a) and (b) are nearly identical, which implies there is no additional loss of coherence with the DQ protocol on this timescale. Given that the contrasts are equal and the period is half for DQ, the maximum slope of the response curve, i.e the sensitivity to magnetic field amplitude, is twice as high for the DQ case. However, for detection of statistical polarization, where the sign of the signal is unknown, it is necessary to measure $B_{rms}^2$, the field power [27]. In this case, it is the *curvature* (i.e the second derivative) of the response curve that determines sensitivity, which is then 4× higher in the DQ case compared to SQ. This implies a potential factor of 4 gain in SNR, with a possible 16× reduction in averaging time for a given SNR. Even in the worst case, i.e. if the DQ coherence time equals half the SQ coherence time, the response to a pulse train of optimal length (of order $T_2$) will be just the same for the SQ and DQ protocols; however, the latter will take half the time. For a given acquisition time, then, the DQ protocol should give at least a $\sqrt{2}$ improvement in SNR [20].

Finally, we have applied the optimized DQ multipulse sequence to detection of real NMR signals. A thin film of the polymer poly(methyl methacrylate), with a proton density of roughly 57 protons/nm$^3$, was spun onto the surface of the diamond. We used optimized 128 pulse sequences for both SQ and DQ detection, and scanned the time $\tau$ between the pulses. The static bias field was 24.1 mT, giving a Larmor frequency for protons of 1.026 MHz, or half period of 487 ns. With 5 minutes of averaging per point, a slight dip was barely detectable in the SQ curve at $\tau = 488$ ns, (Fig. 4(a), depth = $3.0\sigma$, where $\sigma$ is the standard error of the mean determined from repeating the measurements), while for the DQ, the signal was clearly observed (Fig. 4(b), depth = $6.9\sigma$), where the strength of the signal $B_{rms}^2$ is estimated [6-8] to be roughly (90 nT-rms)$^2$. Thus, under these conditions, the SNR was improved by over a factor of 2, which would translate into at least a 4× reduction in data acquisition time. While this performance required careful tuning of the DQ conditions, SNR improvements of at least 1.4 were quite generally obtained.

In summary, the concept of DQ magnetometry has been implemented with a dual frequency pulse scheme and extended to multipulse sequences of up to 128 pulses. The dual frequency approach was shown to be mathematically equivalent to a simple $S_x$ or $S_y$ rotation in one limit. The protocol was applied to a near-surface NV center and used to measure a signal from hydrogen in a polymer film with enhanced SNR.

We wish to thank David Toyli for useful discussions. This work was supported by the DARPA QuASAR program, the Air Force Office of Scientific Research, and the Center for Probing the Nanoscale at Stanford University (NSF grant PHY-0830228).

**Figure Captions**

Fig. 1. (a) The ground state energy levels for the spin-1 nitrogen-vacancy center, showing the zero field splitting $D = 2.87$ GHz and the additional Zeeman splitting of the $m_s = \pm 1$ sublevels. (b) Double quantum dynamic decoupling sequence using a superposition of the $m_s = \pm 1$ states, where the state amplitudes are periodically swapped. The final accumulated phase difference is insensitive to static magnetic fields, but proportional to twice the amplitude of the ac magnetic field (sine wave). Each "swap" is accomplished with a single dual-frequency pulse. (c) (d) Time evolution of the spin wavefunction under dual frequency microwave irradiation of the form $B_{MW}(t) = B_1 \cos(\omega_1 t + \delta_1) + B_2 \cos(\omega_2 t + \delta_2)$, plotted as probabilities, demonstrating the ability to perform a swap operation. In this example, $B_1 = B_2$, and the $m_s = 1$ (−1) state is populated with probability .75 (.25) at $t = 0$ (red square and blue dot, color online). (c) $\delta_1 = \delta_2 = 0°$, (d) $\delta_1 = -\delta_2 = 90°$. In both cases, the wavefunction component probabilities are swapped when $t/t_0 = 1/2$, where $t_0 = 2\pi/\gamma B_1$. "Prepare" and "read" pulses are similarly achieved with $t/t_0 = 1/4$. (e) Circuit for generating the dual frequency pulses [20]. The mixer creates sum and difference frequencies $f_{sum}$ and $f_{diff}$ centered around 2.87 GHz.

Fig. 2. Experimental comparison of (a) Single Quantum (SQ) and (b) Double Quantum (DQ) responses to identical ac magnetic test signals. A simple echo sequence with a single swap was used (echo time = 40 µs). The DQ response oscillates with roughly half

the period, as expected. (c) Echo response *vs* echo time. Solid curve is a fit to the data using the model $\xi(t) = \xi_0 \exp\left(-(t/T_2)^n\right)$. For the SQ (DQ) decay, $T_2 = 82$ μs, $n = 2.2$ ($T_2 = 49$ μs, $n = 1.9$).

Fig. 3. Comparison of (a) SQ and (b) DQ magnetometry using a multipulse sequence of 64 "swap" pulses in response to a magnetic test signal. The repetition time $\tau$ was 528 ns, making the total evolution time 33.8 μs. The test signal was at 947 kHz and was applied synchronously with the pulses (pulse width = 16 ns). (a) was an XY8-64 sequence, while (b) was a related DQ version where the phases were cycled as $\{90°, 0, 90, 0, 0, 90, 0, 90, -90, -180, -90, -180, -180, -90, -180, -90\}$ and repeated four times. The DQ curve has roughly twice the maximum slope and four times the maximum curvature as the SQ curve.

Fig. 4. Detection of NMR signals using multipulse dynamic decoupling sequences with scanned pulse spacing. (a) Signal *vs* pulse spacing for a SQ XY8-128 sequence. The external field was 24.1 mT, corresponding to a proton Larmor frequency $f_L = 1.026$ MHz. A slight dip due to the proton signal is observed at a pulse repetition time of 488 ns, corresponding to $1/2 f_L$ (arrow). Each point is the average of 5 million sequence repetitions (~5 minutes per point). (b) Identical measurement taken with the DQ protocol with phase sequence $\{90°, 0, 90, 0, 0, 90, 0, 90, -90, -180, -90, -180, -180, -90, -180, -90\}$ repeated eight times. A clear enhancement of the SNR is evident.

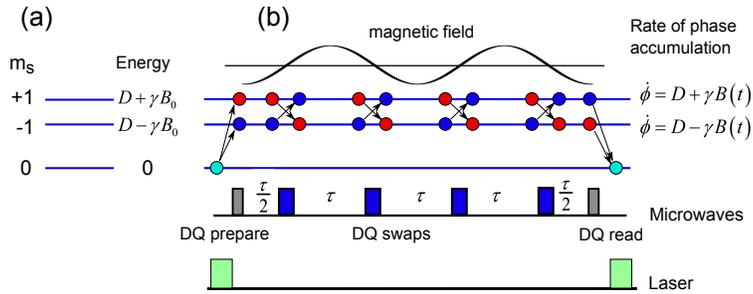

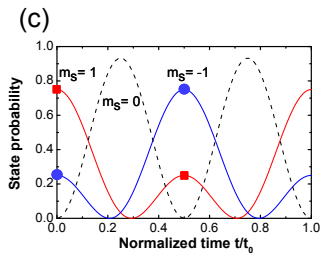 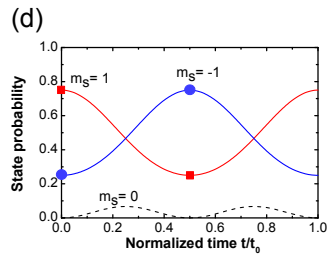

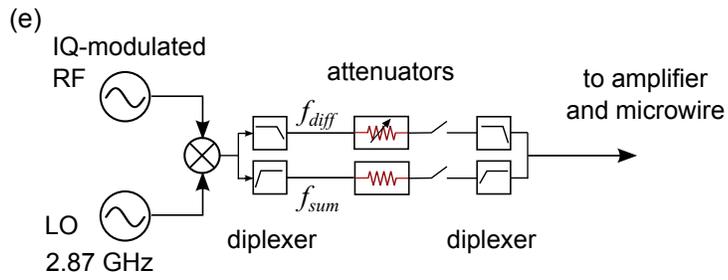

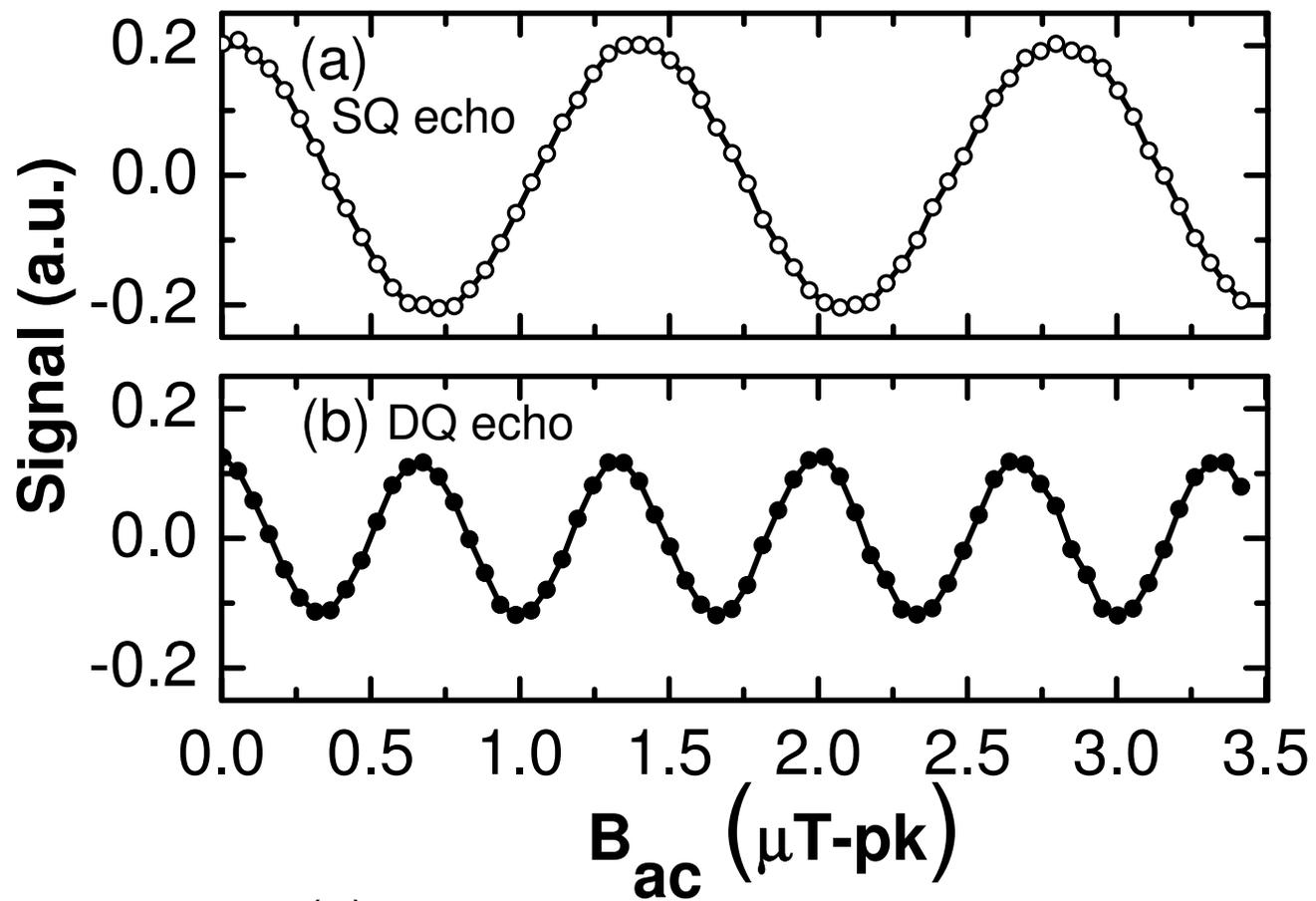
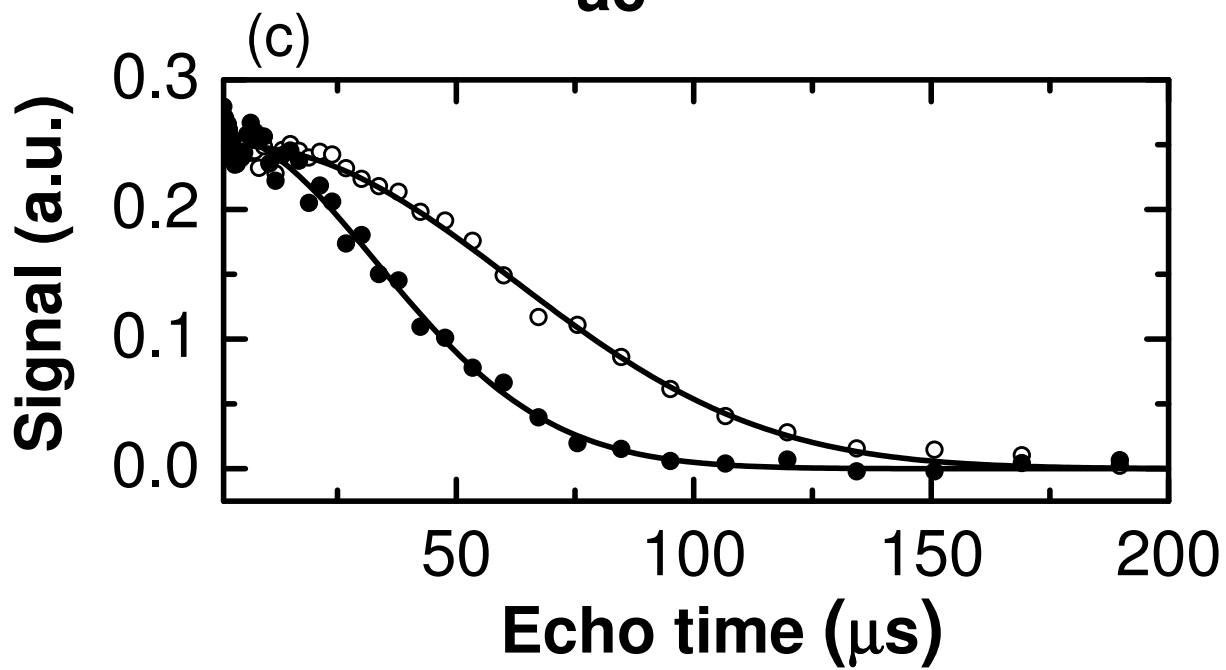

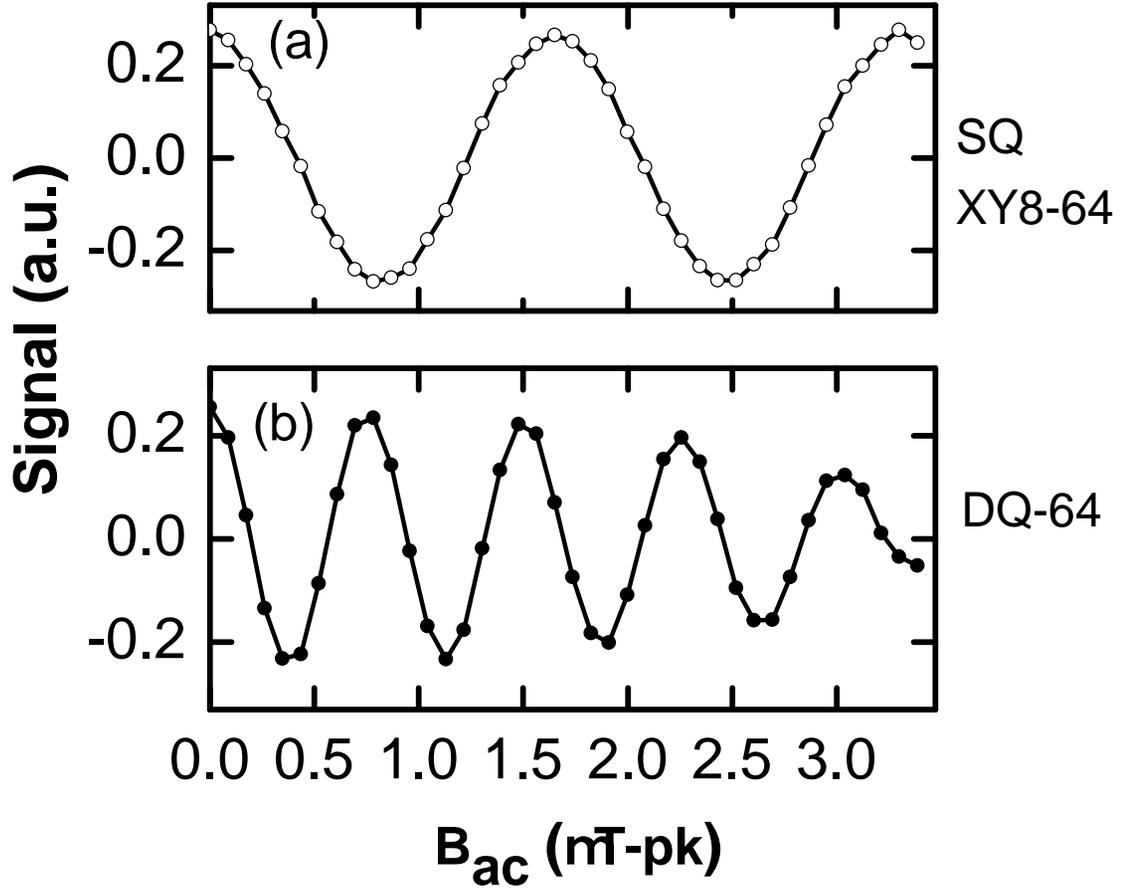

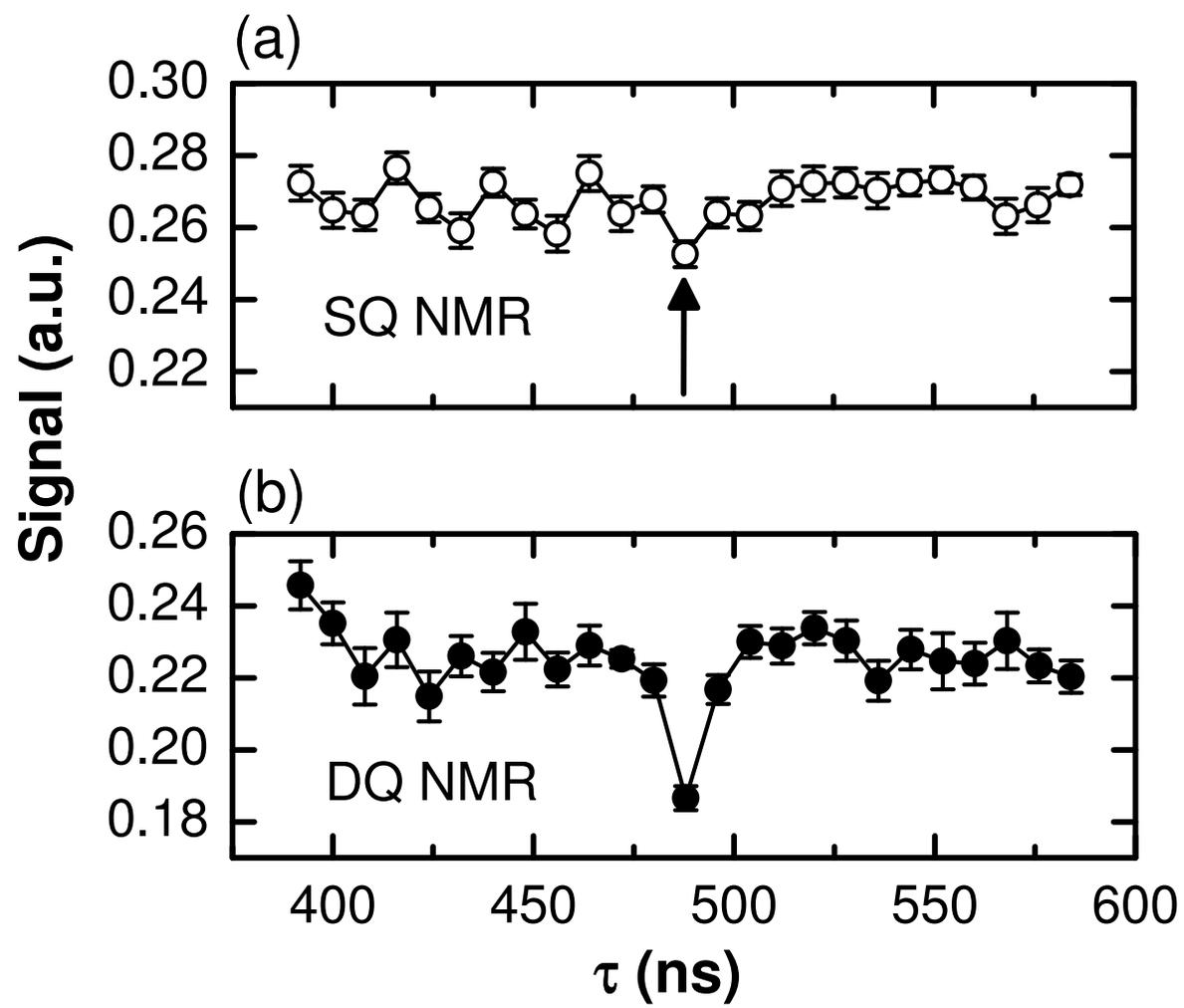

# Multipulse Double-Quantum Magnetometry With Near-Surface Nitrogen Vacancy Centers: Supplementary Material

H. J. Mamin,[1] M. H. Sherwood,[1] M. Kim,[1,2] C.T. Rettner,[1] K. Ohno,[3] D. D. Awschalom,[3,4] and D. Rugar[1]

[1]*IBM Research Division, Almaden Research Center, 650 Harry Rd., San Jose CA 95120*
[2]*Center for Probing the Nanoscale, Stanford University, Stanford, CA 94305*
[3]*Center for Spintronics and Quantum Computation, University of California, Santa Barbara CA 93106*
[4]*Institute for Molecular Engineering, University of Chicago, IL 60637*

## I. Derivation of the Evolution Operator

We start with the Hamiltonian

$H = \gamma B S_z + D S_z^2 + \{\gamma B_1 \cos(\omega_1 t + \delta_1) + \gamma B_2 \cos(\omega_2 t + \delta_2)\} S_x$, which can be written in matrix form in the $m_s = 1, 0, -1$ basis as

$$H = \begin{pmatrix} D + \gamma B & \begin{pmatrix} \gamma B_1 \cos(\omega_1 t + \delta_1) \\ + \gamma B_2 \cos(\omega_2 t + \delta_2) \end{pmatrix} / \sqrt{2} & 0 \\ \begin{pmatrix} \gamma B_1 \cos(\omega_1 t + \delta_1) \\ + \gamma B_2 \cos(\omega_2 t + \delta_2) \end{pmatrix} / \sqrt{2} & 0 & \begin{pmatrix} \gamma B_1 \cos(\omega_1 t + \delta_1) \\ + \gamma B_2 \cos(\omega_2 t + \delta_2) \end{pmatrix} / \sqrt{2} \\ 0 & \begin{pmatrix} \gamma B_1 \cos(\omega_1 t + \delta_1) \\ + \gamma B_2 \cos(\omega_2 t + \delta_2) \end{pmatrix} / \sqrt{2} & D - \gamma B \end{pmatrix}. \quad (S1)$$

We apply the unitary transformation

$$V = \begin{pmatrix} e^{-i\omega_1 t} & 0 & 0 \\ 0 & 1 & 0 \\ 0 & 0 & e^{-i\omega_2 t} \end{pmatrix}$$ to generate the transformed Hamiltonian $\tilde{H} = V^\dagger H V + i(\partial V^\dagger / dt) V$,

which is analogous to transforming to the rotating frame in a spin-1/2 system. The result is

$$\tilde{H} = \begin{pmatrix} D+\gamma B - \omega_1 & e^{i\omega_1 t}\begin{pmatrix} \gamma B_1 \cos(\omega_1 t + \delta_1) \\ +\gamma B_2 \cos(\omega_2 t + \delta_2) \end{pmatrix}/\sqrt{2} & 0 \\ e^{-i\omega_1 t}\begin{pmatrix} \gamma B_1 \cos(\omega_1 t + \delta_1) \\ +\gamma B_2 \cos(\omega_2 t + \delta_2) \end{pmatrix}/\sqrt{2} & 0 & e^{-i\omega_2 t}\begin{pmatrix} \gamma B_1 \cos(\omega_1 t + \delta_1) \\ +\gamma B_2 \cos(\omega_2 t + \delta_2) \end{pmatrix}/\sqrt{2} \\ 0 & e^{i\omega_2 t}\begin{pmatrix} \gamma B_1 \cos(\omega_1 t + \delta_1) \\ +\gamma B_2 \cos(\omega_2 t + \delta_2) \end{pmatrix}/\sqrt{2} & D - \gamma B - \omega_2 \end{pmatrix}. \quad (S2)$$

The off diagonal elements contain both constant and rapidly oscillating terms. For example, the matrix element $\tilde{H}_{01}$ can be expanded as

$$\gamma B_1 e^{i\omega_1 t}\left(e^{i\omega_1 t}e^{i\delta_1} + e^{-i\omega_1 t}e^{-i\delta_1}\right)/2\sqrt{2} + \gamma B_2 e^{i\omega_1 t}\left(e^{i\omega_2 t}e^{i\delta_2} + e^{-i\omega_2 t}e^{-i\delta_2}\right)/2\sqrt{2}$$

$$= \gamma B_1\left(e^{i2\omega_1 t}e^{i\delta_1} + e^{-i\delta_1}\right)/2\sqrt{2} + \gamma B_2\left(e^{i(\omega_2+\omega_1)t}e^{i\delta_2} + e^{-i(\omega_2-\omega_1)t}e^{-i\delta_2}\right)/2\sqrt{2}$$

$$\approx \gamma B_1 e^{-i\delta_1}/2\sqrt{2}$$

where the last step comes from dropping the rapidly oscillating terms (rotating wave approximation.) The result is the transformed Hamiltonian shown in the main text:

$$\tilde{H} = \begin{pmatrix} D + \gamma B - \omega_1 & \dfrac{B_1 e^{-i\delta_1}}{2\sqrt{2}} & 0 \\ \dfrac{B_1 e^{i\delta_1}}{2\sqrt{2}} & 0 & \dfrac{B_2 e^{i\delta_2}}{2\sqrt{2}} \\ 0 & \dfrac{B_2 e^{-i\delta_2}}{2\sqrt{2}} & D - \gamma B - \omega_2 \end{pmatrix}. \quad (S3)$$

The eigenvectors of this matrix represent the eigenstates of the wavefunction in the transformed frame, which we write as $|\tilde{\psi}_i\rangle$, $i = q, r, s$ (the so-called dressed states), each corresponding to its eigenvalue $w_i$.

The wavefunction at any time can be written as a linear combination of the eigenvectors, each of which has a simple time-dependence:

$|\tilde{\psi}(t)\rangle = \sum_{i=q,r,s} d_i \exp(-iw_i t)|\tilde{\psi}_i\rangle$. At $t = 0$,

$|\tilde{\psi}(0)\rangle = \sum_{i=q,r,s} d_i |\tilde{\psi}_i\rangle$, and the coefficients $d_i$ are found via the inner product

$d_i = \langle \tilde{\psi}_i | \tilde{\psi}(0) \rangle$, making use of the fact that the eigenvectors are orthogonal. Thus

$$|\tilde{\psi}(t)\rangle = \sum_{i=q,r,s} \langle \tilde{\psi}_i | \tilde{\psi}(0) \rangle \exp(-iw_i t) |\tilde{\psi}_i\rangle. \tag{S4}$$

If $|\tilde{\psi}(t)\rangle$ is written in matrix form as $\begin{pmatrix} \tilde{c}_+(t) \\ \tilde{c}_0(t) \\ \tilde{c}_-(t) \end{pmatrix}$ in the $m_s = 1, 0, -1$ basis, and each eigenstate is

similarly written as $\begin{pmatrix} e_+ \\ e_0 \\ e_- \end{pmatrix}_i$, then Eq. (S4) can be written as

$$\begin{pmatrix} \tilde{c}_+(t) \\ \tilde{c}_0(t) \\ \tilde{c}_-(t) \end{pmatrix} = (e_+ \ e_0 \ e_-)_q^* \begin{pmatrix} \tilde{c}_+(0) \\ \tilde{c}_0(0) \\ \tilde{c}_-(0) \end{pmatrix} \exp(-iw_q t) \begin{pmatrix} e_+ \\ e_0 \\ e_- \end{pmatrix}_q$$
$$+ (e_+ \ e_0 \ e_-)_r^* \begin{pmatrix} \tilde{c}_+(0) \\ \tilde{c}_0(0) \\ \tilde{c}_-(0) \end{pmatrix} \exp(-iw_r t) \begin{pmatrix} e_+ \\ e_0 \\ e_- \end{pmatrix}_r \tag{S5}$$
$$+ (e_+ \ e_0 \ e_-)_s^* \begin{pmatrix} \tilde{c}_+(0) \\ \tilde{c}_0(0) \\ \tilde{c}_-(0) \end{pmatrix} \exp(-iw_s t) \begin{pmatrix} e_+ \\ e_0 \\ e_- \end{pmatrix}_s .$$

For the on-resonance condition where $\omega_1 = D + \gamma B$, $\omega_2 = D - \gamma B$, the eigenvalues of (S3) are simply $w_r = 0$ and $w_{q,s} = \pm \gamma \sqrt{B_1^2 + B_2^2} / 2\sqrt{2}$. The corresponding normalized eigenstates are found by diagonalizing (S3) to be

$$\begin{pmatrix} e_+ \\ e_0 \\ e_- \end{pmatrix}_{q,s} = \frac{1}{\sqrt{2}} \begin{pmatrix} \frac{B_1 \exp(-i\delta_1)}{\sqrt{B_1^2 + B_2^2}} \\ \pm 1 \\ \frac{B_2 \exp(-i\delta_2)}{\sqrt{B_1^2 + B_2^2}} \end{pmatrix} \text{ and } \begin{pmatrix} e_+ \\ e_0 \\ e_- \end{pmatrix}_r = \begin{pmatrix} \frac{B_2 \exp(-i\delta_1)}{\sqrt{B_1^2 + B_2^2}} \\ 0 \\ \frac{-B_1 \exp(-i\delta_2)}{\sqrt{B_1^2 + B_2^2}} \end{pmatrix}.$$

Plugging these expressions into (S5) eventually gives the expression in the text:

$$\begin{pmatrix} c_+(t) \\ c_0(t) \\ c_-(t) \end{pmatrix} = \begin{pmatrix} \dfrac{B_2^2 + B_1^2 \cos(\omega_e t)}{B_1^2 + B_2^2} & \dfrac{-iB_1 e^{-i\delta_1} \sin(\omega_e t)}{\sqrt{B_1^2 + B_2^2}} & \dfrac{(-1+\cos(\omega_e t)) B_1 B_2 e^{-i(\delta_1-\delta_2)}}{B_1^2 + B_2^2} \\ \dfrac{-iB_1 e^{i\delta_1} \sin(\omega_e t)}{\sqrt{B_1^2 + B_2^2}} & \cos(\omega_e t) & \dfrac{-iB_2 e^{i\delta_2} \sin(\omega_e t)}{\sqrt{B_1^2 + B_2^2}} \\ \dfrac{(-1+\cos(\omega_e t)) B_1 B_2 e^{i(\delta_1-\delta_2)}}{B_1^2 + B_2^2} & \dfrac{-iB_2 e^{-i\delta_2} \sin(\omega_e t)}{\sqrt{B_1^2 + B_2^2}} & \dfrac{B_1^2 + B_2^2 \cos(\omega_e t)}{B_1^2 + B_2^2} \end{pmatrix} \begin{pmatrix} c_+(0) \\ c_0(0) \\ c_-(0) \end{pmatrix}$$

For the general off-resonance condition, we did not find algebraic expressions for the eigenvectors, but for any given time, the eigenvectors could be found numerically and used to generate the evolution operator in a similar way. Finally, to get back to the wavefunction in the lab frame, we apply the transformation $|\psi(t)\rangle = V|\tilde{\psi}(t)\rangle$.

## II. Simulations

One can also numerically solve the Schrodinger equation in the lab frame directly. If

$$|\psi(t)\rangle = \begin{pmatrix} c_+(t) \\ c_0(t) \\ c_-(t) \end{pmatrix},$$ then the time dependence is determined by the Hamiltonian (S1) and

$$i\hbar \begin{pmatrix} \dot{c}_+(t) \\ \dot{c}_0(t) \\ \dot{c}_-(t) \end{pmatrix} = \begin{pmatrix} D + \gamma B & \left(\begin{matrix} \gamma B_1 \cos(\omega_1 t + \delta_1) \\ +\gamma B_2 \cos(\omega_2 t + \delta_2) \end{matrix}\right)/\sqrt{2} & 0 \\ \left(\begin{matrix} \gamma B_1 \cos(\omega_1 t + \delta_1) \\ +\gamma B_2 \cos(\omega_2 t + \delta_2) \end{matrix}\right)/\sqrt{2} & 0 & \left(\begin{matrix} \gamma B_1 \cos(\omega_1 t + \delta_1) \\ +\gamma B_2 \cos(\omega_2 t + \delta_2) \end{matrix}\right)/\sqrt{2} \\ 0 & \left(\begin{matrix} \gamma B_1 \cos(\omega_1 t + \delta_1) \\ +\gamma B_2 \cos(\omega_2 t + \delta_2) \end{matrix}\right)/\sqrt{2} & D - \gamma B \end{pmatrix} \begin{pmatrix} c_+(t) \\ c_0(t) \\ c_-(t) \end{pmatrix}$$

resulting in three coupled differential equations for the coefficients $c_i(t)$:

$$\frac{dc_+}{dt}(t) = -\frac{i}{\hbar}(D+\gamma B)c_+(t) - \frac{i}{\hbar}\frac{(\gamma B_1 \cos(\omega_1 t + \delta_1) + \gamma B_2 \cos(\omega_2 t + \delta_2))}{\sqrt{2}} c_0(t)$$

$$\frac{dc_0}{dt}(t) = -\frac{i}{\hbar}\frac{(\gamma B_1 \cos(\omega_1 t + \delta_1) + \gamma B_2 \cos(\omega_2 t + \delta_2))}{\sqrt{2}} (c_+(t) + c_-(t))$$

$$\frac{dc_-}{dt}(t) = -\frac{i}{\hbar}\frac{(\gamma B_1 \cos(\omega_1 t + \delta_1) + \gamma B_2 \cos(\omega_2 t + \delta_2))}{\sqrt{2}} c_0(t) - \frac{i}{\hbar}(D-\gamma B)c_-(t).$$

These coefficients correspond to the complex amplitudes of the wavefunction in the lab frame, and give the complete time dependence, including that which is analogous to precession about the static magnetic field. This system can be solved numerically using the Runge-Kutta method. This method was implemented in Labview (National Instruments) using a time step of 100 fs.

The probability of finding the NV in each state is then given by $P_i = |c_i(t)|^2$. Figure S1 shows an example of the time dependence of the probabilities for a particular set of conditions (starting in the $|m=0\rangle$ state), as computed both numerically and with the evolution operator from the text. The small oscillations in the numerical method are presumably due to the counter-rotating component of field, which the numerical method includes, unlike the analytic method. Otherwise, the results from the two methods are barely distinguishable. This agreement was seen over the entire range of pulse, frequency and phase conditions that were studied. The chief advantage of the analytic method is that of speed, making it possible to simulate the evolution of the wavefunction very efficiently.

The ability to numerically propagate the wavefunction in time allows us to compute the effect of ac magnetic fields for various multipulse sequences, out to hundreds of pulses or more, in the presence of arbitrary pulse width, frequency and amplitude errors in each microwave component. In general, sequences using the alternating phase sequence $\{0,180°\}$ performed much better than those that used $\{0°\}$ phase through the sequence, due to its compensation against pulse width errors, as described in the text. More generally, however, we found even better results, both in simulation and in practice, with sequences incorporating $\{0, 90°\}$ phase pairs. Figure S2 shows a sample comparison for a 32 pulse sequence applied synchronously with an ac magnetic field, where the pulse widths were all correct, but the effect of a 1 MHz frequency error was considered. We evaluated two different pulse sequences: $(\text{"prepare"})_x - \{(0,180°)\}^{16} - (\text{"read"})_x$, and $(\text{"prepare"})_x - \{(0,90°)\}^{16} - (\text{"read"})_x$. Here the pulses within the $\{\ \}$ are all DQ swap pulses, and the phase $\delta$ as defined in the text is 0° for the prepare and read pulses and alternates between either 0° and 180° or 0° and 90° for the swap pulses.

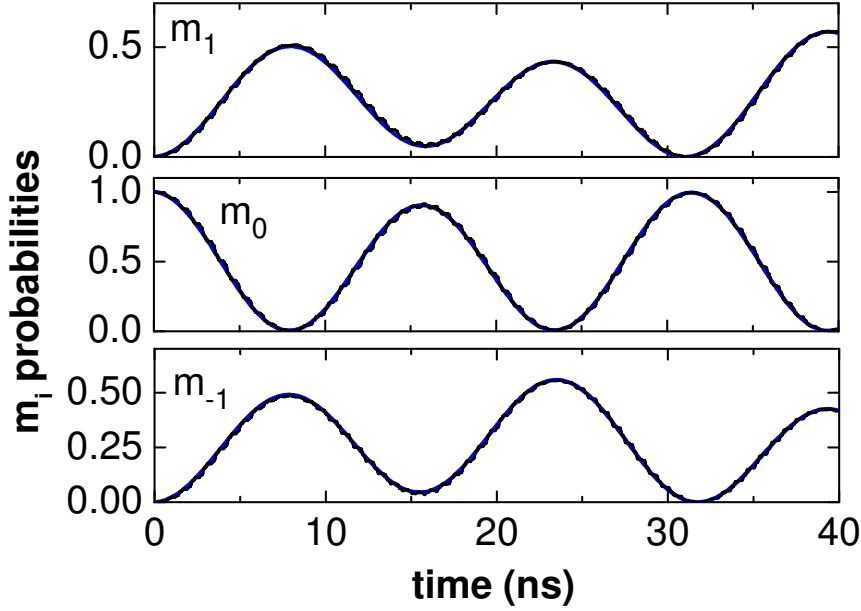

Fig. S1. Time evolution of the wavefunction upon application of the dual-frequency microwave field, plotted as probabilities for the $m_s = +1, 0, -1$ states, with the wavefunction starting in the $m_s = 0$ state. The Runge-Kutta result (black line) and the separately derived evolution operator (blue line) are in nearly perfect agreement. ($B_0 = 25$ mT, $B_1 = B_2 = 2.23$ mT, $f_1 = 3.5$ GHz $= D + \gamma B$, $f_2 = 2.11$ GHz $= D - \gamma B + 10$ MHz (computation assumed $D = 2.8$ GHz).

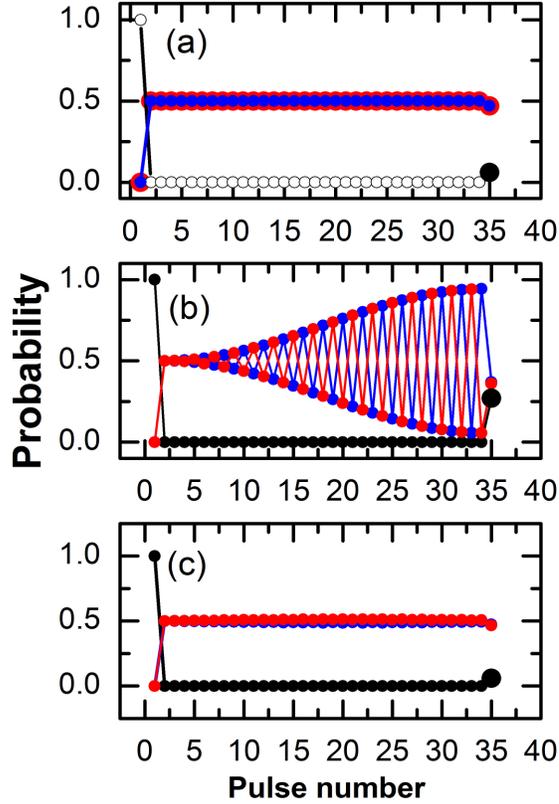

Fig. S2. Simulated evolution during a 32 pulse dynamic decoupling sequence, plotted as state probabilities for the $m_s = +1$ (blue), 0 (black) and -1 (red) states. (Computation assumed $B_0 = 25$ mT, $B_1 = B_2 = 2.232$ mT, swap pulse = 16 ns, prepare pulse = 8 ns, $D = 2.8$ GHz.) (a) On-resonance case ($f_1 = 3.500$ GHz, $f_2 = 2.100$ GHz). (b) Off-resonance case ($f_1 = 3.501$ GHz, $f_2 = 2.101$ GHz: 1 MHz off resonance), with $\delta_1 = -\delta_2 = \{(0,180)\}^{16}$. (c) same off-resonance conditions as (b), except $\delta_1 = -\delta_2 = \{(0,90)\}^{16}$. Note that in (c), the $m_s = \pm 1$ state amplitudes do not show the dramatic deviations from 0.5 that occur in (b), and the final result for the $m_s = 0$ state (black dot) is in much better agreement with the on-resonance case (a).

The plots show the probabilities $|c_i(t)|^2$ during the pulse sequences. When both frequencies are exactly on resonance (Fig. S2(a)), the two sequences give identical results, with the probability of the $m_{\pm 1}$ states equaling ½ after the initial prepare pulse. For this particular value of the applied ac field $B_{ac} = 500$ nT and total evolution time of 15 µs, the final state at the end of the sequence, as shown by the black circles, has $|c_1(t)|^2 \sim 0.07$ for the on-resonance case. When $f_1$ and $f_2$ are both off by 1 MHz, the $\{0,180\}$ sequence shows the $m_{\pm 1}$ probabilities starting to evolve differently (Fig. S2(b)), so that the final state at the end of the sequence is different than the on-resonance case ($|c_1(t)|^2 \sim 0.27$). The $\{0,90°\}$ sequence in Fig. S2(c) shows an evolution much more similar to the on-resonance case, and gives nearly the same final state. This is merely an illustrative example for a particular set of conditions, but sequences based on $\{0,90°\}$ building blocks generally gave better results.

Pulse errors can dramatically affect the sensitivity to ac fields when using multipulse DQ magnetometry. Figure S3 shows experimental data where the response to ac magnetic fields (the change in echo response *vs* field amplitude) is nearly flat (i.e. poor) for small ac fields, which is precisely the region of interest. This example used the alternating $\{0,180\}$ phase sequence. Simulations of the same sequence with deliberately introduced pulses errors (inset) show rather similar qualitative behavior, emphasizing the importance of using robust phase sequences and well-calibrated pulses.

The basic phase sequence used for the NMR detection was the 16 pulse sequence $\{90°, 0, 90, 0, 0, 90, 0, 90, -90, -180, -90, -180, -180, -90, -180, -90\}$, similar to a conventional SQ XY8 sequence followed by another XY8 of opposite phase. This sequence was determined through a combination of simulation and empirical optimization, and was found to be generally robust for small errors in frequency and pulse widths. It was repeated 8 times to form the full 128 pulse sequence.

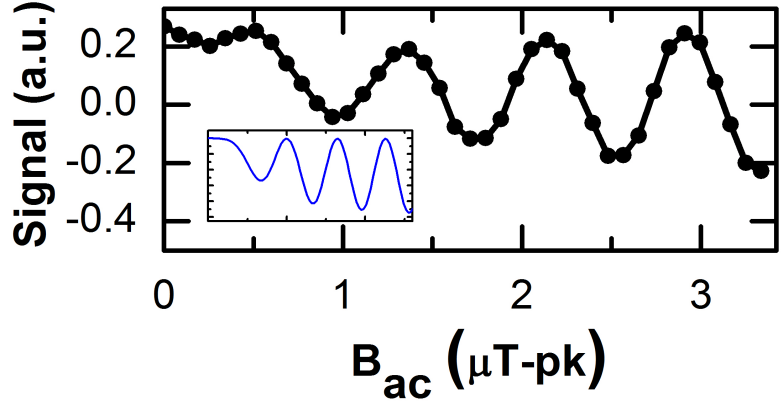

Fig. S3. Experimental multipulse DQ magnetometry using a $\delta_1 = -\delta_2 = \{(0,180)\}^{32}$ sequence similar to that in Fig. S2(b), showing unwanted degradation of the magnetometry response under these conditions. The response to ac magnetic fields is essentially flat for small fields. The inset shows a simulation of the DQ $\{0,180\}^{32}$ sequence where pulse width errors were deliberately introduced, showing that qualitatively similar behavior is possible under some conditions.

**III. Generation of Dual Frequency Pulses**

The use of a mixer provided a relatively simple way to generate dual frequency pulses. A pulsed microwave generator with IQ modulation (Agilent N5182A) was heterodyned with a cw microwave source (HP 83732A) operating at 2.87 GHz using a double-balanced microwave mixer (Marki model M1-0012LQP), as shown in Fig. 1(e). This automatically generates pulses with frequency components at both the sum and difference frequencies, where the pulsed generator frequency was chosen appropriately to match the Zeeman splitting of the lines. In order to balance the strength of the two frequency components, the signal out of the mixer was split into high and low frequency paths with the use of diplexer filters (AMTI Model D2G520G1). A voltage-controlled variable attenuator (RF Lambda model RFVAT0103A30) was then used on the low frequency channel to allow amplitude equalization before recombining the two paths with another diplexer filter. Voltage-controlled microwave switches (RF Lambda models RFSPSTA0003G and RFSPSTA0208G) were inserted into both channels so that each channel could be run separately for calibration purposes. Because of the losses associated with the circuit, primarily with the mixer, we used a 40 dB microwave amplifier at the output of the circuit. The output was then attenuated to the desired level and applied to a gold microwire that had been lithographically defined directly on the diamond.

If the local oscillator input to the mixer has the form $\cos(\omega_{LO} t)$ and the RF input has the form $\cos(\Delta\omega t + \delta)$, then the output of the mixer has product terms of the form

$$\cos(\omega_{LO} t) \cdot \cos(\Delta\omega t + \delta) = \frac{1}{2}\left[\cos((\omega_{LO} + \Delta\omega)t + \delta) + \cos((\omega_{LO} - \Delta\omega)t - \delta)\right].$$

Thus by choosing $\omega_{LO} = D$, the zero field splitting, and $\Delta\omega$ to be half the Zeeman splitting, the output of the mixer has precisely the two desired frequency components. As alluded to in the text, the phase parameters of the two components are equal and opposite with this method.

Balancing the microwave pulse strengths was achieved by performing nutations for each channel individually (i.e. turning one channel off with the switch) and then matching the times for a single quantum pi-pulse. Once this balancing was achieved, both channels were turned on and another nutation measurement was made to determine the proper pulse width and amplitude for a DQ swap. Pulse widths were typically in the 16-32 ns range.

**IV. Diamond Sample Preparation**

We started with electronic-grade single crystal diamond (Element Six) onto which a 64-nm thick isotopically pure layer of $^{12}$C was grown via plasma enhanced chemical vapor deposition [5]. To produce delta-doped NV centers at the desired depth of 12 nm, $^{15}$N$_2$ gas was introduced at the appropriate time during the growth process. The sample was then subjected to $^{12}$C ion implantation to create vacancies [6], followed by an anneal in $2\times10^{-9}$ torr vacuum for 3 hours at $850°$ C, cleaning in equal parts H$_2$SO$_4$, HNO$_3$ and HClO$_4$ at 200°C for 30 minutes, and subsequent heating for 2 hours in oxygen at $425°$ C to remove residual graphitic layers and contaminants.

**V. Signal-to-Noise Ratios for Double Quantum vs Single Quantum Detection in Different Regimes**

The SNR for NV magnetometry is inversely proportional to the minimum detectable field, which has been previously derived for stochastic signals [1–3]. We start with Eq (S14) from Ref. [3] to obtain the relative SNR for DQ compared to SQ:

$$\frac{SNR_{DQ}}{SNR_{SQ}} = \frac{4\tau_{DQ}^{3/2}}{\tau_{SQ}^{3/2}} \frac{\xi_{0,DQ}(\tau_{DQ})}{\xi_{0,SQ}(\tau_{SQ})}. \tag{S1}$$

Here the $\tau$'s refer to the total evolution time used in each protocol, and $\xi_0(\tau)$ is the normalized NV response. Typically $\tau$ is of order of the NV coherence time $T_2$ for optimum performance. The factor of 4 in Eq. (1) comes from the fact that the rate of phase accumulation is twice as fast for the DQ sequence ($\delta\phi_{DQ} = 2\delta\phi_{SQ}$) and that the signal is proportional to the mean square phase $\langle(\delta\phi)^2\rangle$.

In general,

$$\langle(\delta\phi)^2\rangle = \gamma^2 B_{rms}^2 \tau^2 \, h(\tau, T_C), \tag{S2}$$

where the function $h(\tau, T_C)$ accounts for the effects of the finite correlation time of the signal [5]. That is to say, if $B(t)$ is a randomly varying signal with correlation time $T_C$, the accumulated phase will depend on $T_C$. An explicit expression for $h(\tau, T_C)$ has been calculated for both the Hahn echo and a multipulse sequence [5], which allows us to modify Eq. (S1) as follows:

$$\frac{SNR_{DQ}}{SNR_{SQ}} = \frac{4\tau_{DQ}^{3/2}}{\tau_{SQ}^{3/2}} \frac{\xi_{0,DQ}(\tau_{DQ}) h(\tau_{DQ}, T_C)}{\xi_{0,SQ}(\tau_{SQ}) h(\tau_{SQ}, T_C)}. \tag{S3}$$

We now consider the SNR in three distinct regimes:

(A). Suppose that coherence time $T_2$ is the same for both DQ and SQ. Then one can use the same sequence time $\tau_{DQ} = \tau_{SQ}$ (of order $T_2$), so that S2 reduces to

$\frac{SNR_{DQ}}{SNR_{SQ}} = 4$. This is the most favorable limit for DQ.

(B) Suppose that $T_{2,DQ} = T_{2,SQ}/2$, as seen with the Hahn echo. If the sequence time $\tau_{DQ} = \tau_{SQ}/2$ is used, then the same contrast is maintained for the DQ protocol ($\xi_{0,DQ} = \xi_{0,SQ}$). Further assume that the signal correlation time $T_C \gg T_2$, which implies that $h(\tau, T_C) = 1$. In this case,

$$\frac{SNR_{DQ}}{SNR_{SQ}} = \frac{4\tau_{DQ}^{3/2}}{\tau_{SQ}^{3/2}} = 4\frac{(\tau_{SQ}/2)^{3/2}}{\tau_{SQ}^{3/2}} = 2^{1/2}.$$

This is the least favorable regime for DQ. As mentioned in the text, the same SNR is obtained per echo sequence, but the DQ sequence takes half as long.

(C) Consider the case where the magnetic signal has a correlation time $T_C \ll T_2$. In this limit [4],

$h(\tau, T_C) \approx 2T_C / \tau$, and

$$\frac{SNR_{DQ}}{SNR_{SQ}} = \frac{4\tau_{DQ}^{3/2}}{\tau_{SQ}^{3/2}} \frac{\xi_{0,DQ}(\tau_{DQ})}{\xi_{0,SQ}(\tau_{SQ})} \left[ \frac{2T_C / \tau_{DQ}}{2T_C / \tau_{SQ}} \right]$$
$$= \frac{4\tau_{DQ}^{1/2}}{\tau_{SQ}^{1/2}} \frac{\xi_{0,DQ}(\tau_{DQ})}{\xi_{0,SQ}(\tau_{SQ})}.$$

For $T_{2,DQ} = T_{2,SQ} / 2$, this reduces to

$$\frac{SNR_{DQ}}{SNR_{SQ}} = \frac{4}{2^{1/2}} = 2\sqrt{2}.$$

The shorter $T_2$ for DQ relative to SQ does affect the SNR in this regime, but less significantly ($\propto \sqrt{T_2}$), since the signal correlation time is a limiting factor.

Depending on the regime, therefore, the averaging time could be 2×, 8×, or 16× faster. If $T_2$ for the DQ protocol is longer than assumed here, these improvements will become even larger. Of course, this assumes that for multipulse sequences, the effect of pulse errors is no worse for the DQ case as for the SQ case.